\documentclass{aa}

\usepackage{graphicx}
\usepackage{txfonts}
\usepackage{xcolor}
\usepackage{caption}
\usepackage{subcaption}
\usepackage{mathtools}

\usepackage[colorlinks=True,citecolor=blue,linkcolor=blue]{hyperref}
\begin{document}

   \title{Observing gravitational redshift with X-Ray emission in galaxy clusters with Athena X-IFU}


   \author{A. Molin
          \inst{1},
          N. Clerc
          \inst{1},
          E. Pointecouteau
          \inst{1},
          F. Pajot,
          \inst{1},
    E. Cucchetti
          \inst{2}
          }

   \institute{IRAP, Université de Toulouse, CNRS, CNES, UT3-PS, Av. du Colonel Roche 9, 31400, Toulouse, France
             \and
    Centre National d’Etudes Spatiales, Centre spatial de Toulouse, 18 avenue Edouard Belin, 31401 Toulouse Cedex 9, France}


 
  \abstract
   {The Doppler shift predicted by general relativity for light escaping a gravitational potential has been observed on Earth as well as in the direction of various stars and galaxy clusters at optical wavelengths.}
   {Observing the gravitational redshift in the X-ray band within galaxy clusters could provide information on their properties and, in particular, their gravitational potential. We present a feasibility study of such a measurement, using the capabilities of the next-generation European X-ray observatory Athena.}
   {We used a simple generalized Navarro-Frenk-White potential model along with a $\beta$-model for the density of baryonic matter, which sets the emission to provide an estimation of the observed redshift in the simplest of cases. We generated mock observations with the Athena X-ray Integral Field Unit (X-IFU) for a nearby massive cluster, while seeking to recover the gravitational redshift along with other properties of the toy model cluster.}
   {We investigated the observability of the gravitational redshift in an idealized test case of a nearby massive cluster with the Athena
X-IFU instrument, as well as its use in probing the properties of the potential well. We were also able to constrain the mass to a $\sim$20 \% level of precision and the cosmological redshift to less than $\sim$1\%, within a simplified and idealized observational framework. 
   More refined simulations accounting for further effects such as the internal gas motions and the actual shape of the potential well are required to fully investigate the feasibility of measuring the gravitational redshift for a single target or statistically over a sample of galaxy clusters.}
   {}


   \maketitle
%

\section{Introduction}

    \indent Gravitational redshift is caused by the loss of energy of a photon emitted within a gravitational potential and traveling through it. This effect is predicted by general relativity \citep{1916AnP...354..769E} as well as by most alternative gravity theories \citep{cataneo_tests_2018}. The effective associated redshift is given by ${\Delta \Psi}/{c^2}$, where $\Delta \Psi$ is the difference in the gravitational potential between the point of emission and the observer, which is mainly the potential due to the mass of the considered astrophysical object along the line of sight. Hence, the measurement of this redshift can be used to probe either the potential or, equivalently, the mass distribution from which it derives. 

\indent Clusters of galaxies, as the most massive gravitationally bound objects in the Universe, are reasonable candidates for the observation of this effect. Some of the earliest predictions for such observations in clusters of galaxies appear in \citet{cappi_gravitational_1995} and \citet{Broadhurst_2000}. Measurements through optical spectra soon followed, as in \citet{wojtak_gravitational_2011} or, more recently, \citet{2021MNRAS.503..669M} and \citet{2023A&A...669A..29R}. A comprehensive overview is provided in Sect. 4 of \citet{cataneo_tests_2018}, which focuses on tests of gravity with galaxy clusters. In that same section, the authors discuss the observability of the gravitational redshift from X-ray spectra of clusters of galaxies, suggesting that future instruments might be able to achieve such measurements. 

\indent The X-Ray emission from the intracluster medium (ICM) in galaxy clusters arises mainly from the radiative cooling of the hot gas infalling within the halo potential well \citep{1988xrec.book.....S}. The ICM is routinely observed in X-rays from the center of clusters to their outskirts \citep{2019A&A...621A..39E, 2022hxga.book...13W}.
This hot gas is highly ionized and shows strong emission lines from the various elements within it. These emission lines offer access to high precision measurements of the redshift through high resolution spectroscopy \citep{2016Natur.535..117H}.  It is thus  suited for the observation of the gravitational redshift as (at first order), the hot gas distribution follows that of the dark matter, which is the main source of the halo gravitational well. Mapping the weak signal expected from gravitational redshift requires (i) high resolution X-ray spectroscopy in order to retain a high precision over the redshift determination and (ii) a spatial resolution mapping capability to trace the gravitational redshift induced gradient from the center to the cluster's outer parts.

\indent Current X-ray missions such as XMM-Newton or Chandra only provide one of these products at a time, with either low-spectral-resolution imagers such as EPIC \citep{tur01} and ACIS \citep{gar03} or high-spectral-resolution dispersive spectrometers such as RGS \citep{den01}  and LETG/HETG \citep{bri00,can05}. The upcoming generation of X-ray observatory will carry integral field unit spectrometers to offer the capability to achieve spatially resolved high-spectral resolution observation in X-rays. The Resolve instrument \citep{ish22} on board the XRISM missions \citep{tas20} will soon fly, although the observation of the outer parts of clusters will likely be very limited due to the modest size of the XRISM  mirrors impeding the measurement of small redshift gradients. The X-ray Integral Field Unit (X-IFU, hereafter)  on board the Athena observatory implements the science theme of the "hot and
energetic Universe" \citep{2013arXiv1306.2307N} and it should provide the adequate performances. 
The X-IFU is required to have a  5~arcmin field of view (FoV) with a full width half maximum (FWHM) resolution of 5~arcseconds and a  spectral resolution of 2.5~eV over the 0.2-7~keV energy range \citep{2018SPIE10699E..1GB, bar23}. 
With this work, we investigate the feasibility of measuring the gravitational redshift in massive clusters of galaxies with the X-IFU instrument.

\indent The work and results presented in this paper were obtained with the current baseline configuration for the Athena mission. Because of the actual programmatic context, the European Space Agency is revisiting the formulation of the Athena mission science case and specifications. Our results may thus be affected by to the to-be-defined instrumental configuration of the Athena mission.
Throughout this study, we assume a $\Lambda$CDM cosmology with  $h=H_0/100\,\textrm{km/s/Mpc}=0.7$, $\Omega_\Lambda = 0.7$ and $\Omega_m = 0.3$. In this framework, at a redshift of $z=0.1$, 10 kilo-parsecs (kpc) correspond to a angular extent of 5.4~arcsec. 

\section{X-IFU/Athena mock observations}

\indent In order to investigate the observability of the gravitational redshift from the X-Ray emission of galaxy clusters, we used a cluster toy model, based on the simulations presented in \citet{2018A&A...620A.173C}, to produce simulated observations with Athena  X-IFU using the SIXTE instrument end-to-end simulator \citep{wilms2014athena, 2019A&A...630A..66D}. The emission models and spectral fitting rely on the \texttt{xspec} software \citep{1996ASPC..101...17A}. 

\subsection{The X-IFU instrument}
As a next-generation European X-ray observatory, Athena \citep{bar17}, will board an integral field unit spectrometer with unprecedented capabilities, the X-IFU.  It will allow for the spatial mapping of emission lines over extended sources such as galaxy clusters, allowing for spatially resolved spectroscopy with a power of R$\sim$1000 \citep{2022cosp...44.2316B}. 
 The X-IFU will be equipped with a high precision detection chain including an array of more than a thousand Transition Edge Sensors (TES) cooled to 55~mK and high precision  readouts electronics. It will provide the required  high-spectral-resolution of 2.5~eV FWHM over the 0.2-7~keV energy band. Combined with the large collective area of the Athena mirrors, it will benefit from an effective area of $\sim$1$\mathrm{m}^2$ at 1\,keV. The requirement for the spatial resolution of the Athena mirrors is 5~arcsec half energy width (HEW). Taken together, these performances will fully open the era of spatially resolved high spectral resolution at X-ray wavelengths, in the wake of the first glimpses provided by the SXS instrument onboard the Hitomi satellite and of the upcoming observation of the Resolve instrument \citep{2023arXiv230301642S} on board the XRISM mission \citep{2022arXiv220205399X}.

\subsection{Cluster toy model}

\indent For the purpose of our study, we chose to model a nearby massive cluster, with $z=0.1$ and $M_\textrm{200}=10^{15}$~M$_\odot$. 
Accounting for the faintness of the gravitational redshift effect, local and very massive clusters are ideal targets to aim at its detection. Lower and/or more distant clusters would render such detection almost impossible and, as such, they are not further considered in this study.
A more detailed discussion on the cluster choice is provided in Sect. \ref{gravzsim}. The parameters of the cluster according to the model described below are summarized in Table~\ref{cluster}. The angular size of the cluster at this distance, noted $\theta_{200}$, is provided as well. 

The cluster toy model consists of a gas density model and a dark matter density model. The cluster is discretized as a grid of emitting particles, to which the parameters of the emission model are assigned based on their position in the cluster. The size of the grid is chosen such that it contains one X-IFU FoV and is deep enough to contain ${R_{200}}$ of the cluster along the line of sight. At the chosen redshift, this corresponds to a grid of 7500 kpc in depth (i.e., along the line of sight), and 938x938 kpc in width.

\subsubsection{Redshift}

The redshift of photons emitted in the cluster is the composition of multiple sources, which are detailed with the following equations from \citet{cataneo_tests_2018}, for the emission point $\vec{x}$ and an observer lying at the origin of the reference frame: 
\begin{equation}
    1+z_{\mathrm{tot}} = (1+z_{\mathrm{cosmo}})\left[1+\frac{1}{c^2}(\Psi(0) - \Psi(\vec{x}))+\frac{\vec{n}\cdot\vec{v}}{c} + \frac{v^2}{2c^2}\right]
\label{e:red}
\end{equation}
where $z_{cosmo}$ is the cosmological redshift, $\Psi$ is the gravitational potential, $\vec{n}$ is the unitary vector parallel to the line of sight, and $\vec{v}$ is the velocity vector of the emitting point relative to the observer. The two last terms correspond respectively to the Doppler shift along the line of sight and the relativistic transverse Doppler shift. In the ICM, these are mainly due to the bulk and turbulent motions of the gas. We deliberately chose not to address these intrinsic motions of the gas in our study (we further discuss this choice in Sect.~\ref{s:dis}). The resulting approximation is then:
\begin{equation}
    z_{\mathrm{grav}} = \frac{\Delta \Psi}{c^2}
.\end{equation}
 
\subsubsection{Dark matter density model}
\label{DMmodel}

\indent We assumed that the dark matter (DM) density follows a generalized Navarro Frank White radial profile (hereafter, gNFW). The gNFW profile has been worked out based on a generalization of the NFW profile \citep{1997ApJ...490..493N, 2007ApJ...668....1N}. The gNFW profile has three slope indexes, $\alpha$, $\beta,$ and $\gamma$, where $\beta$ is the inner slope and $\gamma$ is the outer one. We used a version presented in \citet{2019arXiv190105615Z}, which sets $\alpha$ and $\beta$ to 1. The profile is otherwise characterized by $r_s$, a scale radius, the overdensity, $\delta_c$, and $\gamma$ is the asymptotic slope when $r\xrightarrow{}0$. The scale radius, $r_s$, is related to  the mass, $M_{\delta}$, at the density contrast, $\delta,$ (different from the overdensity) times the critical density of the Universe at redshift $z$, $\rho_{\text{crit}}(z)$, as follows:
   
   \begin{equation}
      r_s = \left( \frac{M_{\delta}}{\frac{4}{3} \pi \delta \rho_{\text{crit}}(z)}\right)^\frac{1}{3} \frac{1}{c_\delta}
   ,\end{equation}
   with $c_\delta$ being the concentration parameter. The overdensity, $\delta_c$, can be expressed as a function of $M_{\delta}$  as follows :
   \begin{equation}
       \delta_c = \frac{M_{\delta}}{\int_0^{R_{\delta}} \frac{4 \pi r^2 \rho_{\text{crit}}(z)}{(r/r_s)^\gamma (1+r/r_s)^{3-\gamma}}}
   ,\end{equation}
with $R_{\delta} = c_\delta \cdot r_s$. This expression can be developed in the case of a gNFW density profile, as provided in the Appendix \ref{AppedixA}. This entirely describes the DM density from which the gravitational potential can be derived analytically \citep{2019arXiv190105615Z}. In doing this, we neglect the contribution of the gas and stars to the gravitational potential. We use this model in the following sections. We also chose to add a constant to the potential $\phi$ to set $\phi(r\xrightarrow{}+\infty)=0$. This allows for a straightforward conversion between the potential and the redshift of light emitted from a point $\textbf{r}$ in the cluster such that $z = {\Psi(\textbf{r})}/{c^2}$ or, when expressed as an equivalent veolcity shift, $v_z =  {\Psi(\textbf{r})}/{c}$. 

\subsubsection{Gas density model}
\label{emissionmodel}
We modeled the emission of our toy model cluster ICM with a broadened APEC model (\texttt{bapec}) under \texttt{xspec} \citep{2001ApJ...556L..91S}. This model represents the emission of a collisional, optically thin, diffuse plasma, mainly through the Bremsstrahlung radiation for the continuum, as well as the atomic lines due to the different processes at play in the plasma (e.g., dielectronic recombination, ionization, and radiative transitions). The broadening of the lines is only thermal in our simulations, excluding other possible sources of broadening such as bulk motions or turbulence. 


For this study, we restrained ourselves to a simple isothermal cluster with homogeneous abundance through the cluster. We set the temperature such that $k_B T = 7$ keV. The solar abundances follows that of \citet{1989GeCoA..53..197A} and we set the intra-cluster gas global abundance such that $Z/Z_{\mathrm{solar}} = 0.7$. This leaves only the redshift and the normalisation as varying parameters for the \texttt{bapec} model. 

The norm of each emitting volume element, V, of the cluster is defined as:
\begin{equation}
    N = \frac{10^{-14}}{4 \pi(D_A(1+z))^2} \int n_e n_p dV
,\end{equation}
with $D_A$ as the angular distance of the cluster in cm, $z$ as the cosmological redshift, $n_e$ and $n_p$ as the electron and proton particle densities in $\mathrm{cm}^{-3}$, respectively. The resulting norm is given in photons per unit of volume, per unit of effective area, thus: $\text{cm}^5$. For  a fully ionized plasma, we can consider $n_e = 1.2 n_p$. The emission model is multiplied by a photo-absorption model, \texttt{phabs} under XSPEC, using cross-sections from  \citet{1996ApJ...465..487V}, to account for the Galactic absorption. We fixed the hydrogen column density, noted $n_H$, to $0.03 \times 10^{22} \text{cm}^{-2}$. \\  
\indent For analytical convenience, we adopted a simple $\beta$-model \citep{1976A&A....49..137C} as our gas density model, although it is not the best fit to represent the actual distribution of the intra-cluster gas. It is parameterized by the core electron density, $n_0$, the core radius, $r_c$, and the slope, $\beta$. 

\subsection{Foreground and background emissions} 

\indent We accounted for the astrophysical foreground and background emissions in our simulations following the model proposed by \citet{2002ApJ...576..188M}. It includes a non absorbed thermal model representing the local bubble (\texttt{apec}), a second absorbed one for the Galactic halo (\texttt{phabs*apec}), and an absorbed power law for the cosmic X-ray background (CXB, \texttt{phabs*powerlaw}). We adopted the parametrisation provided by \citet{2014A&A...569A..54L}.
The hydrogen column density is kept at the same value as for the cluster model. 

\indent The instrumental background is also accounted for in our simulations.  It is managed entirely by the SIXTE tool according to the X-IFU requirements of $5\times10^{-3} [\mathrm{counts/s/cm}^2\mathrm{/keV}]$. This instrumental background mainly results from the high-energy cosmic rays hitting the neighborhood of the detector. 





\subsection{Observational strategies}
\label{obs_strat}

\begin{figure}[!h]
    \includegraphics[width = 9cm]{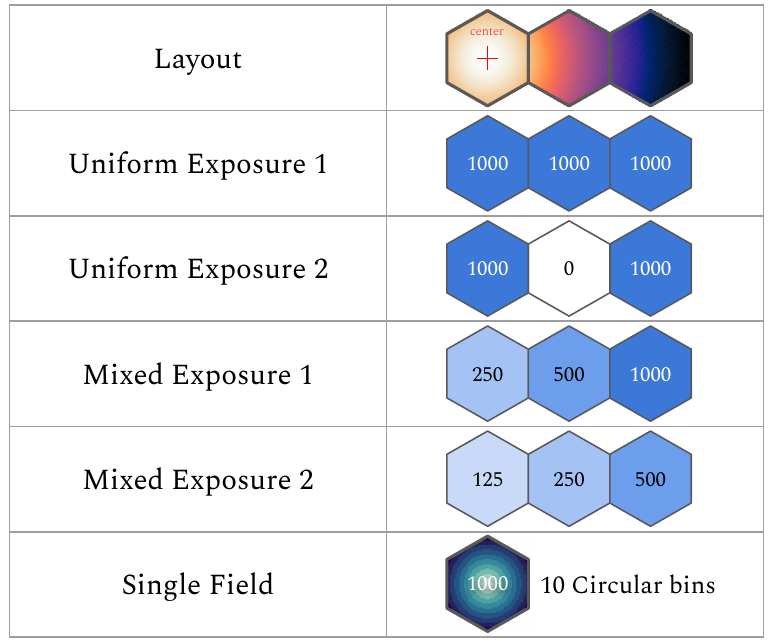}
   \caption{Observing strategies considered in our simulations. The layout of the X-IFU pointings is shown in the right column together with the exposure time for each pointing in ksec. The count map for the configuration "uniform exposure 1" is plotted in the first line and shows the cluster center.}
   \label{summarytable}%
\end{figure}

\indent We investigated various observational configurations in order to assess the feasibility of measuring the gravitational redshift with the X-IFU instrument on board Athena. We varied the number of X-IFU pointings from one to three and individual exposures from 125~ksec (kiloseconds)  to 1~Msec. The six investigated configurations are illustrated in Fig~\ref{summarytable}. The various multiple pointings configurations allow us to sample measurements of the ICM emission as far as the characteristic radius of   $\sim0.6 R_{200}$ ($\sim0.9 R_{500}$).

\begin{table}
   \centering
      \caption[]{Parameters of our toy model clusters for the gravitational potential, DM and gas densities, and the gas emission.}
         \label{cluster}
      \begin{tabular}{c c} 
            \hline
            Parameter  &  Value \\
            \hline
            $M_{200}$ & $1\cdot10^{15} M_\odot$  \\
            $R_{200} (\theta_{200})$ & 2 Mpc (18.5')   \\
            $c_{200}$ & 4.5      \\
            $\gamma$ & 1.2 \\
            $z_{\mathrm{cosmo}}$ & 0.1 \\
            $r_c$ & 400 kpc  \\
            $\beta$ & 2/3 \\
            $n_0$ & $3\cdot10^{-3}$ [$\mathrm{cm}^{-3}$]\\
            $k_B T$ & 7 [keV]\\
            Abundance ($Z/Z_\odot$) & 0.7 \\
            nH  & 0.03 [$10^{22}\text{cm}^{-2}]$ \\
            \noalign{\smallskip}
            \hline

        \end{tabular}
\end{table}

\begin{figure*}[!h]
   \includegraphics[width = 18cm]{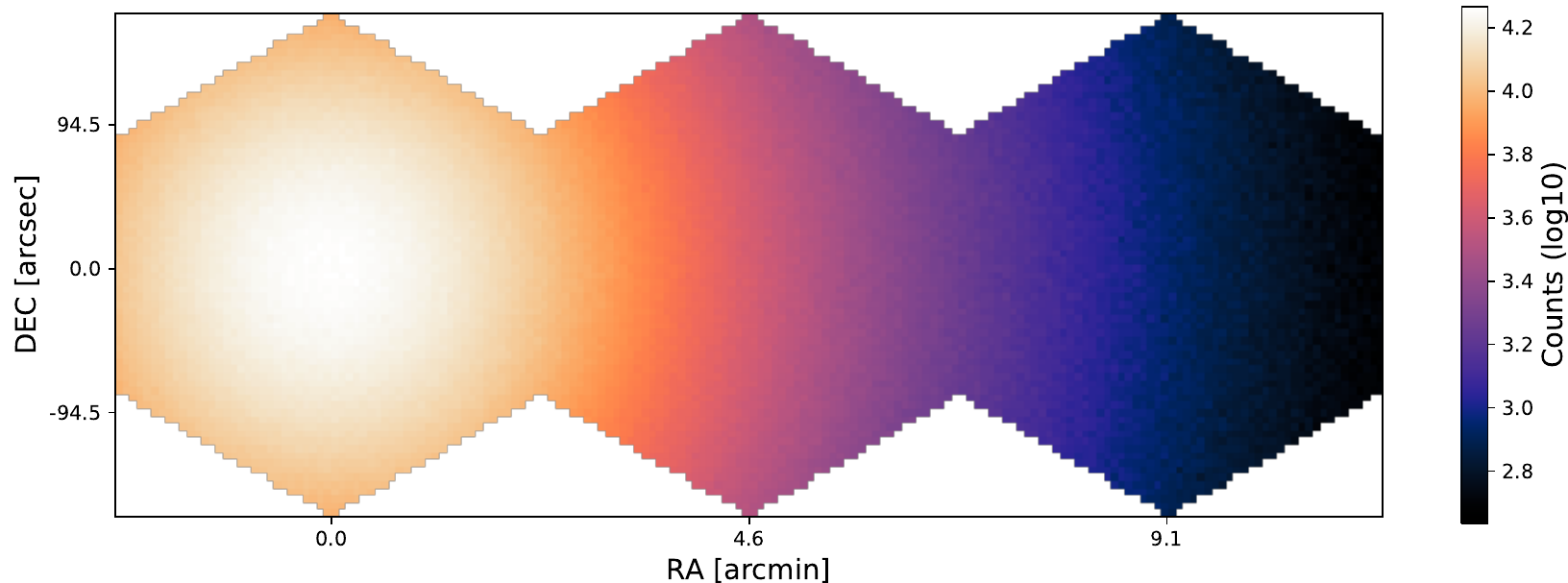}
   \caption{Count map in each pixel ($\sim$5x5~arcsec) of three adjacent 1Ms pointings of X-IFU (corresponding to uniform exposure 1 in Table \ref{summarytable}) of a $10^{15}M_\odot$ and $z=0.1$ cluster. The color scale is in units of $\log_{10}$ of the counts. The center of the cluster is at (RA, DEC) = (0, 0)}
              \label{countmapbin}%
\end{figure*}

\section{Mock data analysis}
\label{s:dataanalysis}

\indent The main output of the SIXTE simulator is a mock event list of the X-IFU observation. For all the recorded events, namely, the X-ray photons that have been detected, the measured energy, detector and sky coordinates, time of arrival, and so on, are provided. From the SIXTE mock event lists, we generated count images. The spectra were computed within concentric annuli of constant width or over  full X-IFU FoV. Each spectrum is fitted using \texttt{xspec} with the \texttt{phabs*bapec} model and the aforementioned model for the background, with the redshift, the cluster emission normalisation, temperature, abundance, and the background emission normalisations as free parameters. The velocity broadening of the \texttt{bapec} model, which is set to 0 when simulating the cluster emission, is also set to 0 during the fit. The justification for this choice is discussed in Sect.~\ref{s:line}. From the best-fit value in each bin, we reconstructed the redshift profile used to assess the gravitational redshift.
In Figure \ref{countmapbin}, we show an example of a count map for an observation of three contiguous pointings  (i.e., the configuration named "uniform exposure 1," as defined in Figure~\ref{summarytable}).


\subsection{Modeling the observed gravitational redshift} \label{gravzsim}

\indent From the DM potential well and gas emission models described in Sects.~\ref{DMmodel} and \ref{emissionmodel}, we modeled the  observed redshift as an emission-weighted redshift along the line of sight. Then, it is expressed as:  

\begin{equation}
    z_{obs, los}(\vec{\theta}) = \frac{\int_l z( \textbf{r}) \epsilon (\textbf{r}) dl}{\int_l \epsilon ( \textbf{r}) dl} 
,\end{equation}
with $\epsilon( \vec{r} )$ as the emissivity at $\vec{r}$, $l$ as the line of sight, and $\vec{\theta}$ as the angular distance to the cluster center. The finite dimension of the grid for the cluster model restricts the precision of the integrals to a finite length. We quantified that the loss in flux due to this cutoff is less than 1\% by computing the integrals of the emissivity for different cutoff values. We approximate the integrals with a double exponential quadrature integral \citep{takahasi1974double}, which allows for a very good approximation (within numerical errors) and important computational time improvement. The redshift in a single bin is obtained with :
\begin{equation}
    z_{bin} \ =\ \frac{\int _{s_{bin}} z_{obs,los}( \vec{\theta} ) \epsilon _{obs,los}(\vec{\theta} ) d\vec{\theta} }{\int_{S_{bin}} \epsilon _{obs,\ los} d\vec{\theta} } 
,\end{equation}
with $S_{bin}$ the area of the bin and $\epsilon _{obs,\ los} = \int_l \epsilon ( \textbf{r}) dl$. This formula remains true for any bin shape.

\begin{figure}[!h]
   \includegraphics[width = 9cm]{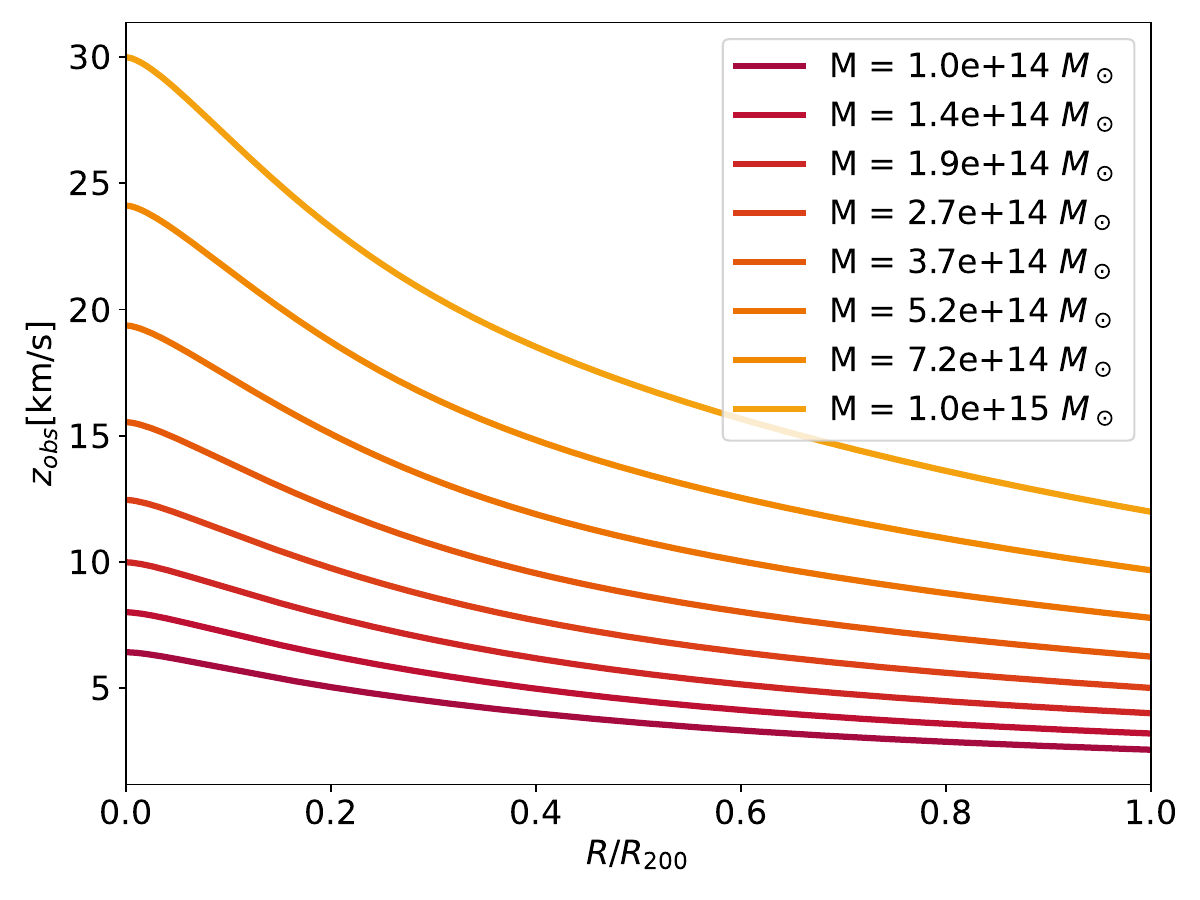}
   \caption{Emission weighted radial profiles of the gravitational redshift for clusters of different masses.}
              \label{masses}%
\end{figure}

\indent Models of the observed scaled radial profile for the gravitational redshift, $z_{obs}$, are shown on Fig.~\ref{masses} as a function of the cluster mass. 
As expected, more massive clusters show a deeper and steeper potential, making them obvious target for measurements of the gravitational redshift.
The drawback is the angular size and extent of the cluster in view of the X-IFU FoV, limiting the emission sampling in the outer parts of clusters in a single X-IFU pointing. 
Even when the emission of more distant (hence, less extended) clusters at a given mass would be better sampled spatially, it quickly suffers from the dimming of the X-ray flux with redshift. The need for a balance between apparent luminosity and angular size, led us to choose a local massive cluster as a test case for our study.

\subsection{Line shift measurements and fitting procedure}
\label{s:line}

\indent Figure~\ref{masses} shows how the measurement of the gravitational redshift in a nearby massive cluster requires to measure redshifts with a precision of a few km/s. This is almost an order of magnitude lower than the line shift expected from bulk motions and turbulence in the ICM \citep{2022hxga.book...56K, 2019SSRv..215...24S}.
The requirement on the line shift and line broadening precisions for the X-IFU in its current configuration is respectively of 10 and 20~km/s ($1\sigma$ level), for a typical observation time of $\sim$100~ksec. 
This imposes the need for an energy scale precision to better than 0.4 eV at $\sim 6$~keV ($1\sigma$ level) and set over the 0.2-7~keV energy range \citep[e.g., ${0.4\,\text{eV}/}{6\,\text{keV}}\cdot c \simeq 20 \text{km/s}$ at $\sim$6~keV][]{10.1117/12.2312170}). This means that no incoming photon can have its energy determined with a precision better than 0.4 eV. It should not, however, be interpreted as a strict limitation on line energy and, thus, the line speed measurements.
Over a whole observation, the factors leading to the variation of the energy scale will be corrected every few ksec (currently 4ks considered for the X-IFU), and are expected to vary evenly around the 0 point. 
This means that over a typical observation time (i.e., 10-1000~ksec), the energy scale variations should mainly result in a broadening of the lines. Assuming that other instrumental systematics are under control, the uncertainty on the line shift will be the only one remaining. It should remain below 10~km/s for a 100~ksec exposure time observation. It may thus be neglected in our 1~Msec simulations. We also note that the current version of SIXTE does not implement the effect of this in-flight energy scale correction.

For a given spectral resolution, given enough time, any precision over a Gaussian line centroid can be achieved. The associated error on the line centroid, $\sigma_v$, goes as $\sigma_v \sim {\sigma_{res}}/{\mathrm{S/N}}$ (with S/N as the signal-to-noise ratio, typically the square root of the number of counts for photon noise). Hence, the only restriction for measuring line shifts, for a given resolution, is observing time, or conversely, for a given observation time, is the instrumental resolution. The approximation over the centroid given previously is true for a Gaussian line \citep[see][for extended details]{cucchetti2019astrophysique}. 


\indent Besides the physical motivation of introducing thermal broadening, the reason for choosing the \texttt{bapec} model over a simple \texttt{apec} model is practical.  The main contributing element for the redshift measurement in the fitting procedure is the lines. In the case of a simple APEC model, the lines are considered as infinitely thin. This means that fitting the line position is automatically limited by the energy bin width of the instrument's response. More precisely, the likelihood becomes discretized and usual fitting procedures, such as gradient descent, do not guarantee proper convergence and/or proper parameter error estimation. We expand on this issue and illustrate it with a plot in Appendix \ref{linefitting}. However, with broadened lines (as is the case for a \texttt{bapec} model), this problem does not arise, and we can be confident in the fitting procedure and its outcome for the redshift. Hence, we used a \texttt{bapec} model for modeling and fitting the emission of the cluster. The velocity broadening was set to 0 in both cases, leaving only the thermal broadening accounted for. 

\indent Our spectra are binned according to the \citet{2016A&A...587A.151K} method. They are then fitted over the whole X-IFU energy band, namely, 0.2-12 keV. We  chose to make use of the whole information carried by the many emission lines of the ICM. We considered that this maximizes the use for the lines signal to extract the redshift. In the case of a real cluster (or a more evolved model), this would also allow for enough precision to be provided on the redshift with varying density, temperature, and abundances depending on the regions of the cluster investigated (i.e., central parts or  outskirts). A focus on a single specific strong line (e.g., Fe~K-$\alpha$) could also deliver a constrained enough estimation of the gravitational redshift. However, such a investigation is beyond the scope of this paper.



\section{Results}

\begin{figure}[!h]
    \includegraphics[width = 9cm]{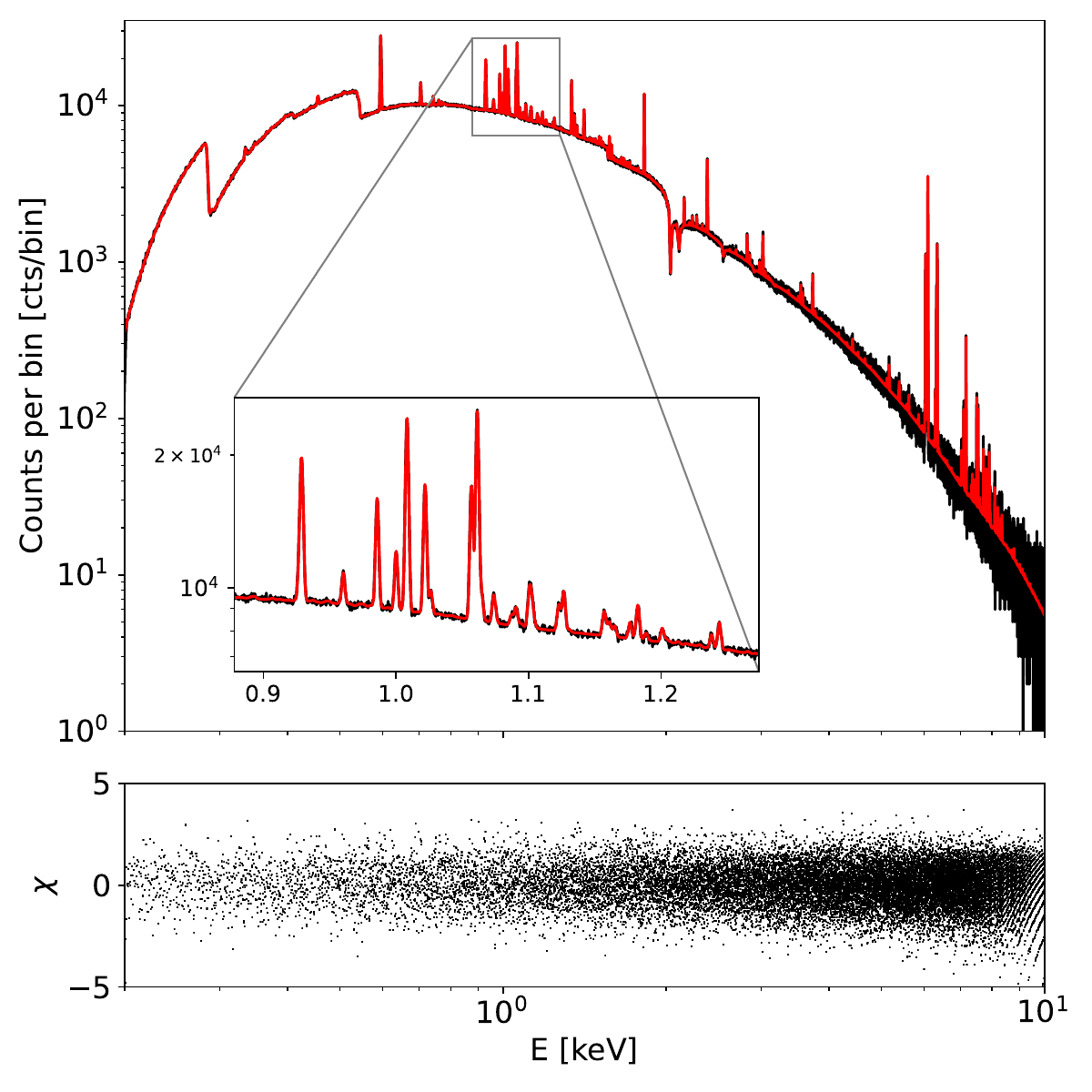}
   \caption{Simulated spectrum from a 1Ms observation of the center of our cluster toy model (black line) and its best fit (red line). This spectrum is extracted from the central bin of the circular binning shown in the summary table in Fig. \ref{summarytable}. An inset figure shows the details of the lines observed around 1 keV. The lower panel shows the error in units of $\chi$ for each bin. The quantized shape observed at the highest energies comes from the low number counts observed at these high energies.}
   \label{spec}%
\end{figure}

\indent In this section, we first evaluate the precision in the construction of the redshift profile that is achievable for a simple, single X-IFU 1Ms pointed observation of our toy model cluster. This allows for an evaluation of the reproducibility of such a measurement. In the second part, we investigate other observational strategies, using multiple X-IFU pointings and exposures to evaluate whether the observed redshift profile can be used as a probe to constrain parameters of the gravitational potential. We show an example the spectrum obtained with a 1Ms observation of the center of the cluster in Figure \ref{spec}. 

\subsection{Recovery of the gravitational redshift}
\label{s:res}

\begin{figure}[!t]
    \includegraphics[width = 9cm]{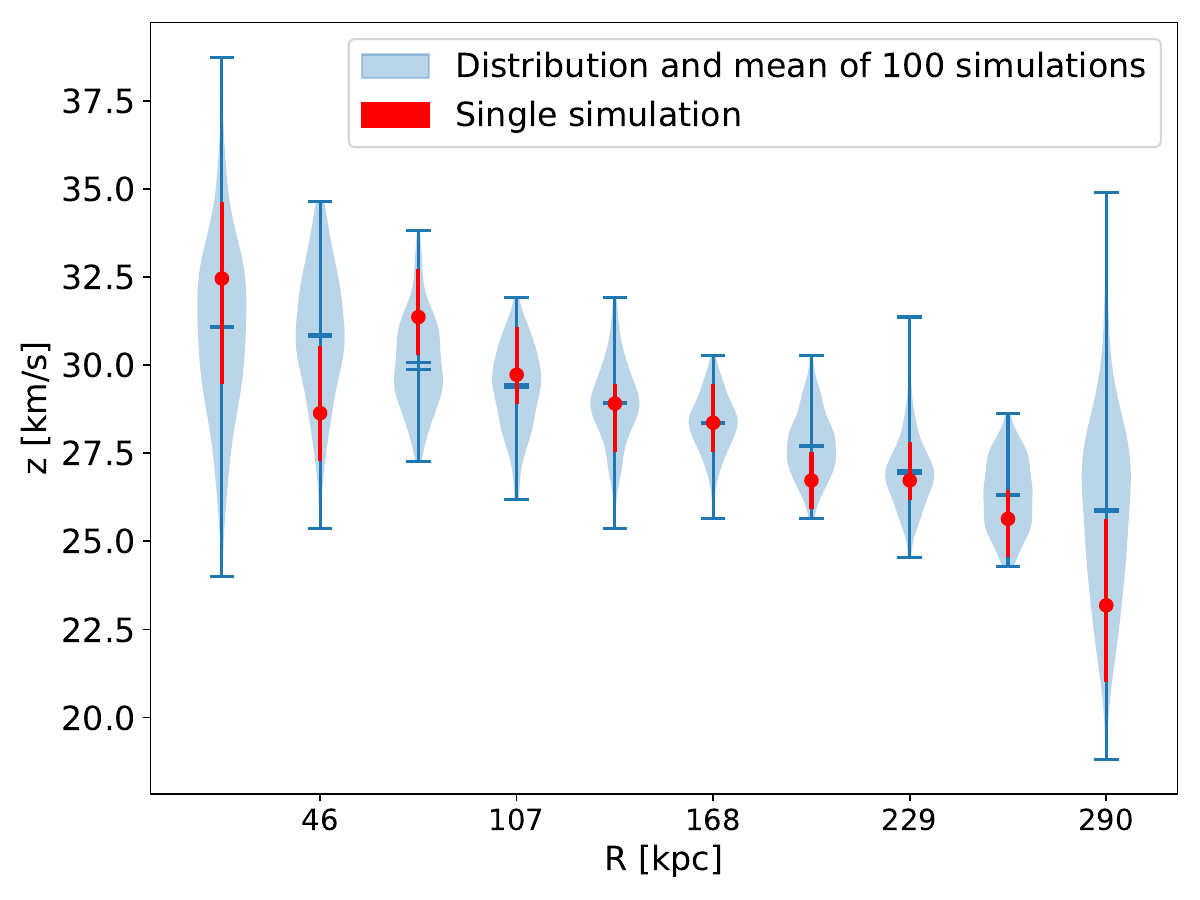}
   \caption{Reproductibility in the reconstruction of the radial profile of the gravitational redshift over a single X-IFU pointing at the center of our toy model cluster. The blue points and associated errors show the mean profile and its associated dispersion over a 100 simulations of a 1Ms observation (see Sect.~\ref{s:res}.) The red points show the example of a single profile and the associated errors provided by \texttt{xspec}.}
   \label{zprofilenobkg}%
\end{figure}

\indent From the various observing configurations of single or multiple X-IFU pointings as defined in Sect.~\ref{obs_strat},  we are able to retrieve the radial profile for the redshift. The Poisson noise in our simulations is the only stochastic process. To check the reproducibility of our reconstructed redshift profile with respect to this source of noise, we ran 100 simulations for our "single-field" observational configuration. 
Figure~\ref{zprofilenobkg} shows the mean profile and its dispersion over the 100 reconstructed profiles, together with a single profile with its error bars derived from the \texttt{xspec} fit. The profiles are reconstructed over ten circular concentric annuli from the cluster center and covering the whole FoV. The conversion to velocities shown on the y-axis assumes a  prior knowledge of the cosmological redshift ($z_{\mathrm{cosmo}}$ in Eq.~\ref{e:red}). This exercise of reproducibility illustrates the dispersion of profiles, which is not fully encapsulated within the error bar of each single measurement. This exercise has been led only for this specific configuration because of its heavy computational demand (about 2~hours for a single simulation on 32 cores CPUs, hence, a total of about 200~hours for the reproducibility study on a single toy model cluster).  However, changing the binning and/or exposure time should not affect the recovery of the profile on average. 

\subsection{Constraining the cluster parameters}

\begin{figure}[!h]
    \includegraphics[width = 9cm]{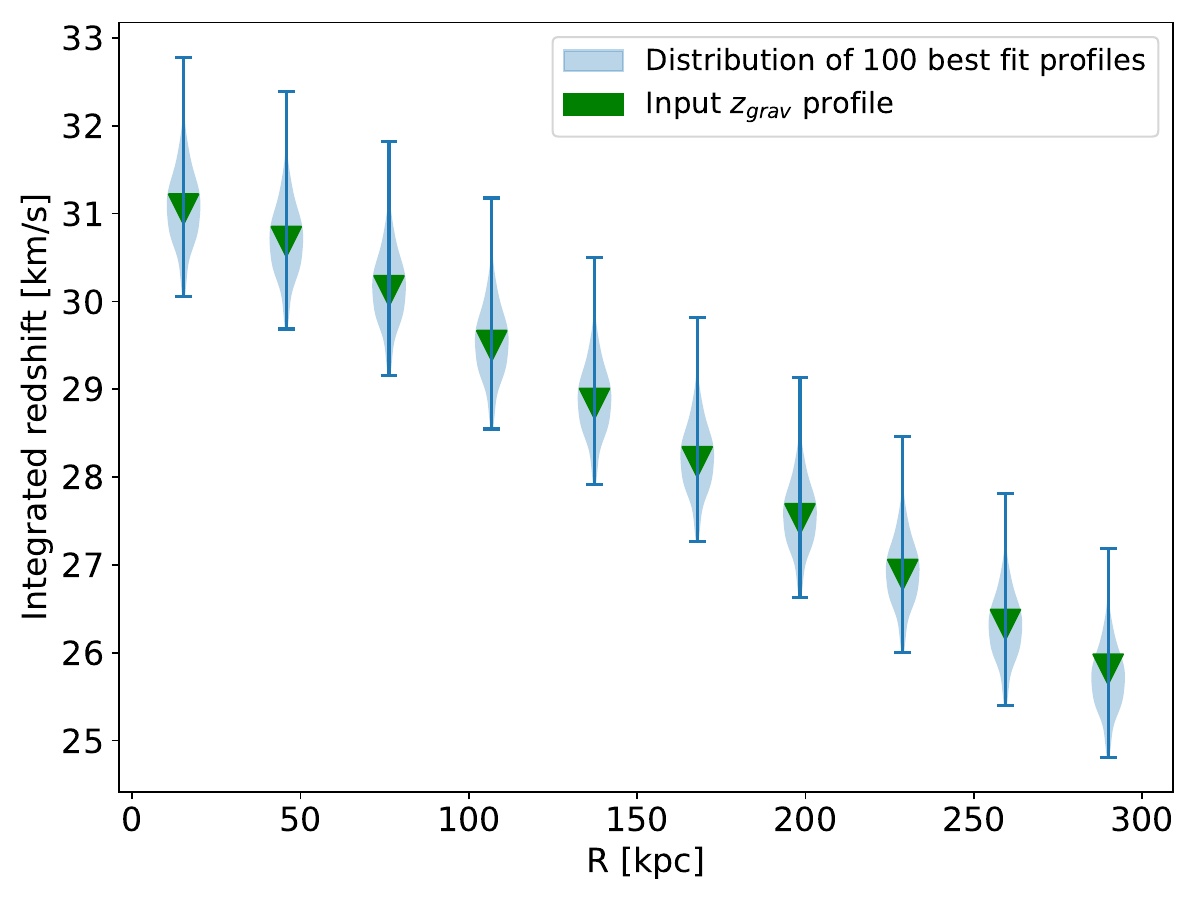}
   \caption{Distribution of the best fit profiles to 100 simulated reconstructed  gravitational redshift profiles from X-IFU mock observations . The expected profile, that was used as an input in all the simulations, is plotted with the green triangles.}
   \label{bestfitsprofiles}%
\end{figure}

\indent The gravitational redshift directly links to the halo potential well and, thereby, to the underlying total mass of the cluster. The measurement of the gravitational redshift profile can thus be used as a probe to determine the cluster mass. The assumption of a prior and perfect knowledge of the cosmological redshift allows for one of the parameters of the model to be fit, such as the cluster mass. We can fit the expected redshift profile from the model detailed in Sect.~\ref{gravzsim} to the redshift profile obtained from our mock observations. As a toy model, we used a simple least squares minimization for the fitting procedure. Figure \ref{bestfitsprofiles} shows the distribution of the 100 best-fit curves, fitting only the mass of the cluster. The distribution of the best-fit profiles is centered on the expected input profile, showing little to no bias in the profile recovery. This idealistic situation leads to an exceedingly optimistic estimation of the halo mass. We obtained a mean best-fit mass of $0.998 \pm 0.018 \cdot 10^{15} M_\odot$ (for an expected mass of $10^{15} M_\odot$).


\indent In reality, the situation would be less optimistic as none of the cluster parameters (e.g., the density, temperature, shape of the DM distribution, etc.) would be known perfectly. As such, they will have to be determined from the X-ray observations or constrained from ancillary data. The cosmological redshift could for instance be constrained from optical observations. The precision of this redshift would then condition our ability to estimate the gravitational redshift. All these uncertainties and unknowns would have to be formulated as priors in our analysis.
As a first step towards this more complex situation, we considered the cosmological redshift to be completely undetermined.
Because the mass and cosmological redshift  similarly impact the observed redshift profile, we need to use the entire shape of the profile to disentangle their correlated effect (see, e.g., Fig. \ref{masses}). 
 To address this issue, we used multiple pointings mock observations (see Fig.~\ref{summarytable}).
To maximize the signal-to-noise ratio (S/N) in the determination of the redshift, we considered each pointing of our three pointings configurations as a single bin and derived the associated spectra over the whole X-IFU FoV. In the case of a real potential well, small-scale variations in the gravitational field (and, thus, in the gravitational redshift distribution) could be expected, although this is not the case for our smooth gNFW toy model.   


\begin{figure*}[h]
    \begin{subfigure}{0.5\textwidth}
        \centering
        \includegraphics[width=\textwidth]{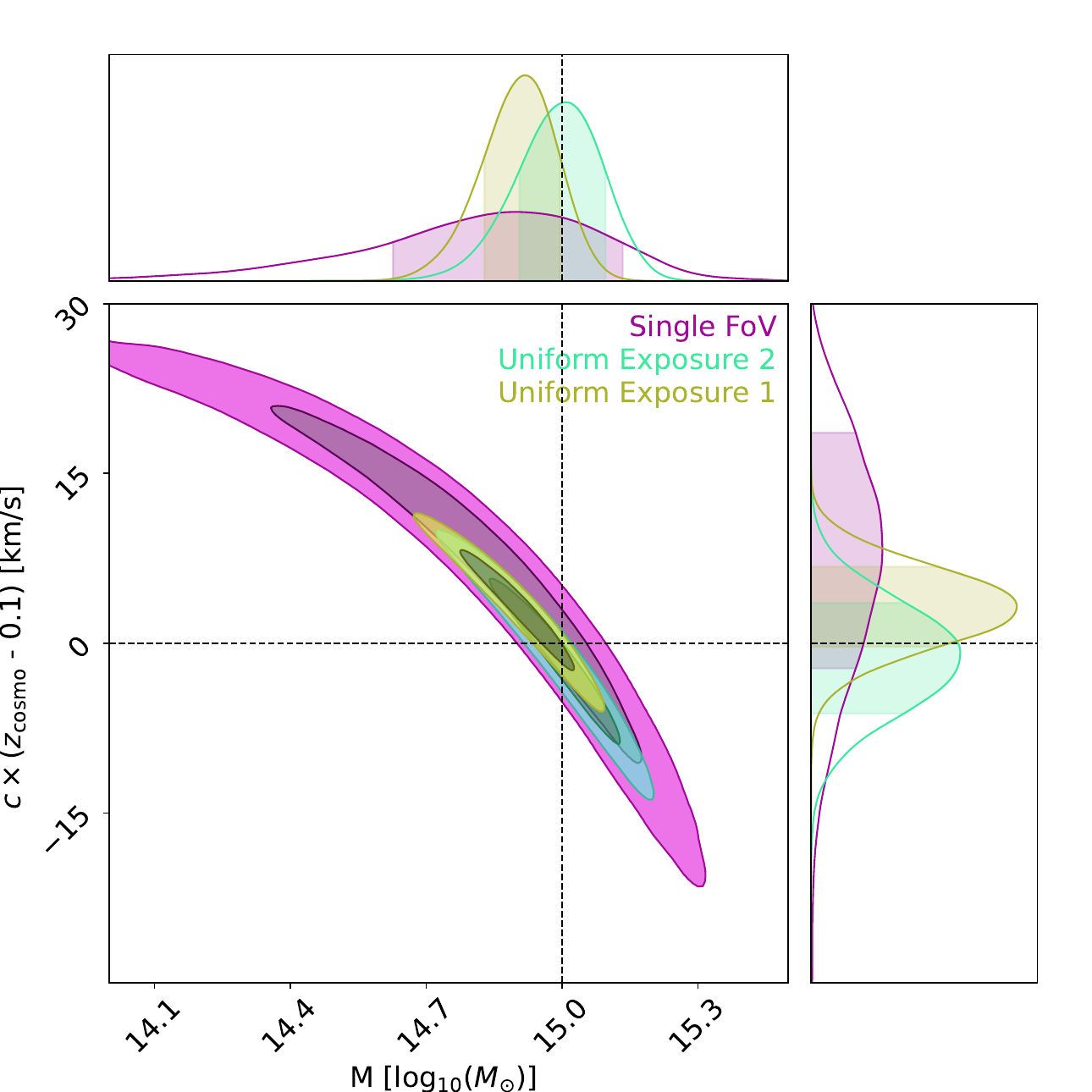}
        \label{uniformbkg}
    \end{subfigure}
    \hfill
    \begin{subfigure}{0.5\textwidth}
        \centering
        \includegraphics[width=\textwidth]{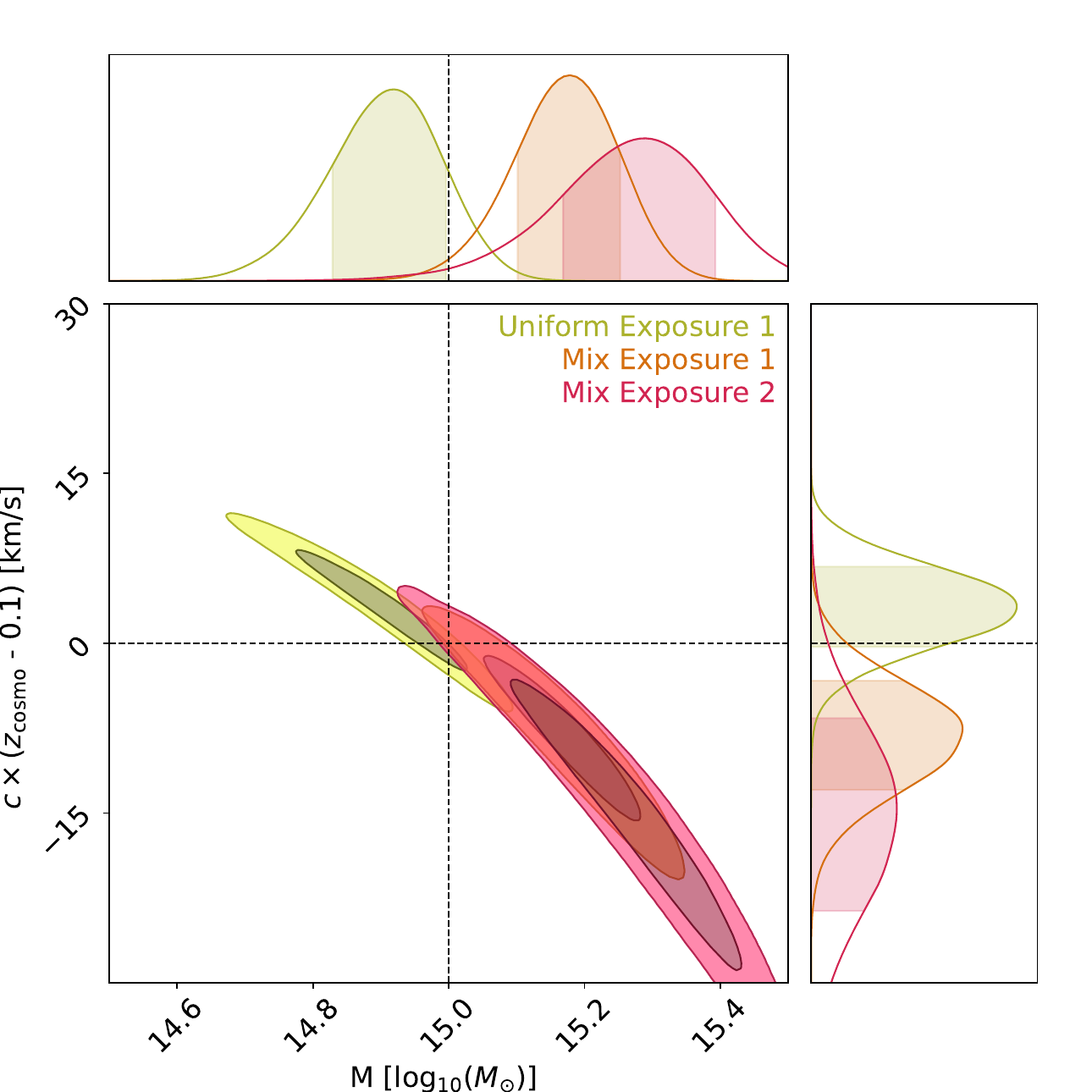}
        \label{mixbkg}
    \end{subfigure}
    
    \caption{Corner plots in the mass and redshift plane and associated posterior distribution for both parameters. The different colors correspond to the different observing strategies listed in the legend and presented in Fig.~\ref{summarytable}. Left:\ Uniform exposures 1 and 2, compared with a single pointing observation. Right: Mix exposures 1 and 2, compared with a single pointing observation. The input reference values are plotted as dotted lines.}
    \label{cornerplots}
\end{figure*}


In the left panel of Figure \ref{cornerplots}, we compare the posterior distribution for mass and redshift obtained with the Markov chain Monte Carlo (MCMC) method (using the \texttt{emcee} package) on three different observing configurations, with a single, two and three X-IFU pointings, respectively. The single pointing configuration is binned with ten circular annuli, whereas the two multiple pointings configurations are binned into a single region for each pointing. For the best case scenario, that is "uniform exposure 1," the constraint obtained on the mass is $ M = 0.80^{+0.17}_{-0.14} \cdot 10^{15} M_\odot$, whereas the single field alone provides $ M = 0.63^{+0.58}_{-0.30} \cdot 10^{15} M_\odot$. The errors are provided at the 68 \% confidence level. {While "uniform exposure 2"} seems to be more centered on the true value, the first one brings more constraints and, thus, lower errors on the reconstruction of the mass and redshift due to the added third pointing (see Fig.~\ref{summarytable}). The shift with respect to the true values of the parameters is due to the sample variance, which affect both configurations similarly, each being a single statistical realisation of the cluster emission (see the reproducibility study in Sect.~\ref{s:res}). We evaluated the Pearson correlation coefficient obtained over the mass and redshift samples in the distributions for each of these strategies. In all of them, the coefficient remains above 0.9 in absolute value. The strong degeneracy between the mass and redshift is only restrained to a smaller range for strategies mapping the outer parts of the cluster. In the presented case, the astrophysical and instrument backgrounds have little impact, as we have set relatively important exposure times and since we are targeting a nearby massive cluster (see Appendix~\ref{appendixnobkg} for further details on the simulations without background). 

\indent We also performed simulations with more realistic exposure times. In order to  optimize the S/N across the radial range probed by our multiple pointings observations, we doubled the exposure time from one pointing to the next adjacent one (see Fig~\ref{summarytable}: "mixed exposure 1" accounts for a  total of 1.75~Msec exposure, whereas "mixed exposure 2" for a total of  875~ksec. The results for these observing  strategies are presented in the right panel of Fig.~\ref{cornerplots}). "Mixed exposure 1" provides $ M = 1.48^{+0.28}_{-0.23} \cdot 10^{15} M_\odot$ and "mixed exposure 2" provides $ M = 1.84^{+0.56}_{-0.43} \cdot 10^{15} M_\odot$. 

\indent The posterior distribution retrieved from mixed exposures 1 and  2 in Fig.~\ref{cornerplots} are centered more than 1$\sigma$ away from the input values. We believe that the line of sight mixing causes an underestimation of the errors that cannot be accounted for by the \texttt{bapec} model under \texttt{xspec}. Because the likelihood used for the MCMC is using these very same errors from \texttt{xspec}, the retrieved parameter distribution is showing optimistic error levels. In addition,  the mixed exposure 1  strategy gives a heavier weight to the pointings far from the center. This compensates the lower signal in these regions. However, the fit in these regions can be biased not only from the line of sight mixing but also from the stronger contribution of the background; hence, this encompasses a biased redshift measurement and a biased posterior distribution.

\begin{figure}
   \includegraphics[width = 9cm]{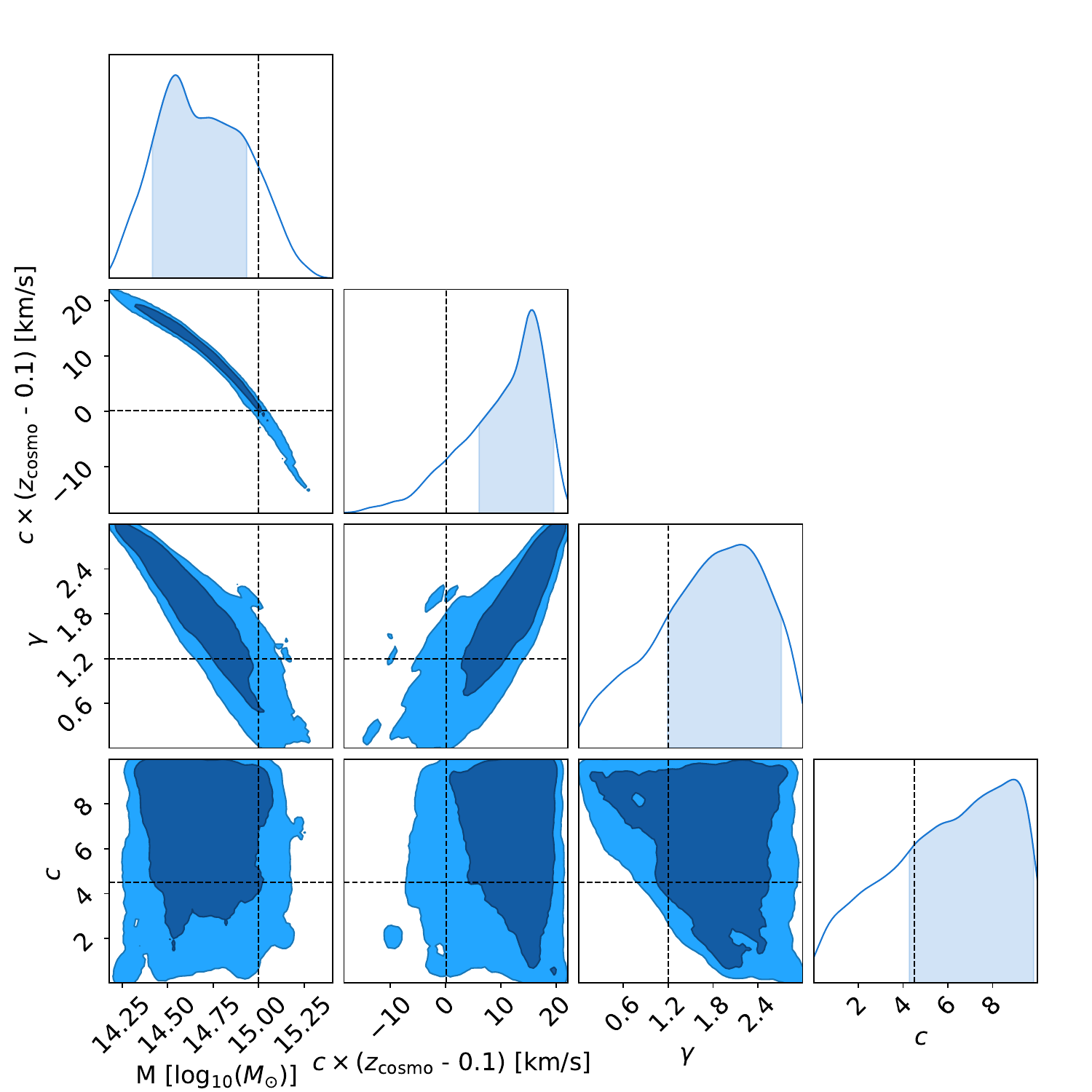}
   \caption{Corner plot of the mass, redshift, $\gamma$ and $c_{200}$ gNFW parameters retrieved from the MCMC fit of the  redshift profiles derived  from the observing configuration named "uniform exposure 1." The true values, used as input to the simulations, are plotted with dotted lines.}
   \label{allpar}%
\end{figure}

\indent The previous tests assume that the shape of the potential was known, namely, that the parameters $\gamma$ and $c_{200}$ were fixed at their known input values. A final test was run freeing all the parameters of the gravitational potential,  including the mass, $\gamma,$ and $c$. The result is shown in Figure \ref{allpar}. The distributions show that all the parameters span a large range of values, typical of unconstrained models. The concentration parameter, $c$ is especially poorly constrained; it spans the entire uniform prior range, from 0 to 10. The upper left distribution shows the strong correlation between the mass and the cosmological redshift when the shape of the expected redshift profile is not fixed by the other shape parameters. This shows that the gravitational redshift alone cannot constrain all the parameters of the potential. We recall that it is highly unlikely that such measurements would be carried out from the X-ray point of view only, without any other ancillary data sets or inputs (e.g., gravitational lensing).  

\section{Conclusions and discussion}
\label{s:dis}



\indent In this work, we evaluated the possibility of observing the gravitational redshift in galaxy clusters in X-rays with future integral field spectrometers such as the Athena X-IFU. To that end, we created mock observations of an idealized massive and nearby galaxy cluster (the targets with the highest probability to be detected) with X-IFU, by using the SIXTE software. 
We analyzed the data with the \texttt{xspec} spectral analysis software. We reconstructed the  gravitational redshift profile that we modeled through the shape of the cluster  potential well and the X-ray emission of its gas content. We  showed that: (1) X-IFU could recover the gravitational redshift for massive ($M_{200} \sim 10^{15}$~M$_\odot$) and nearby ($z \sim 0.1$) clusters within a quite large, but still achievable  exposure time; and (2) the measurement if the gravitational redshift profile can be used to derive  properties of the halo gravitational potential, such as its total mass.

\indent These conclusions have to take into account the limitations of our model. Firstly, we stress that the gas mass fraction in our simulated cluster is relatively high ($\sim$20 \% for $M_{\mathrm{gas}}/M_{500}$ at $R_{500}$). This is due to our choice of a $\beta$-model for the gas distribution, which can overestimate the gas fraction at large radii. Moreover, our choice of total mass, that is $M_{200} = 10^{15} M_\odot$, is rather conservative with respect to some local massive clusters \citep{2019A&A...621A..39E}. Assuming $M_{200} \simeq 1.5 \cdot 10^{15} M_\odot$ would yield a gas fraction of about 15~\%. With such a cluster, we would observe a higher amplitude for the redshift profile. The mass does not affect the S/N of the observed X-Ray spectra. Hence, this does not change the essence of our results, as the uncertainties come from the photon counts, and these are driven by the emission, exposure and distance.  

\indent  Secondly, we neglected the motions of the gas in the cluster. These motions result in a Doppler shift in the emission, which is of the order of $\sim$100-1000~km/s \citep{2022hxga.book...56K, 2019SSRv..215...24S}. This is an order of magnitude above the observed redshift. It means that in a real cluster, the observation of the gravitational redshift would be added to that of bulk motion. The gravitational redshift would then be a difficult quantity to estimate. However, we can consider things the other way round, and any precise measurement of bulk or turbulent velocities using line shifts will have account for the gravitational redshift as a systematic bias. An a priori knowledge of the total mass profile, thus of the gravitational potential would provide the proper estimate of such a bias on bulk and turbulent motions of the ICM hot gas.
In addition to turbulence, the internal structures of the physical properties of galaxy clusters (density, temperature, pressure, abundances, etc) depart at various scales from the idealized  hypothesis of sphericity and homogeneity we adopted \citep[e.g.,][]{2012ARA&A..50..353K, 2022hxga.book...65L}. 

\indent The line of sight mixing is another issue when challenging the limits of spectral precision. Because the ICM is optically thin, we observe the emission of all the points along the line of sight. This increases the signal, but causes the different emitted spectra to be mixed. Because many of the observed quantities are not additive and we are modeling the observed spectrum with a single model, 
we assumed that the observed profile, which is an emission weighted average over different redshifts, is the profile of the emission weighted average redshift. This assumption works because the center of the cluster is the most emitting part and thus dominates the signal. However, this would not be systematically the case for a non-spherically symmetric potential and/or ICM emission. This obviously would also concern other physical parameters such as the temperature and the chemical abundance. 
The line of sight mixing remains a weak effect, as we illustrate in Fig. \ref{spec}, where the observed spectrum is perfectly overlapping with the model. The evaluation of the fit of all our mock observations holds a ${\chi^2}/{ \mathrm{d.o.f} }$ in the 1 to 1.15 range. 

\indent One way around these issues could be to stack several observations of different clusters. In doing that, fluctuations from cluster to cluster, such as the shape and the turbulence, could average out, and the gravitational redshift would remain. This would then require scaling the clusters with respect to each other, as well as other practical considerations, such as the determination of the cluster center. Such investigations  have already been done with optical data, and have been used to test alternative theories of gravity \citep{wojtak_gravitational_2011, 2021MNRAS.503..669M, 2022arXiv220605313R}. Similar work could be undertaken on the basis of observations of clusters samples with future X-ray Integral Field Units, such as X-IFU.

\indent By the time Athena X-IFU is launched, exploratory work could be carried out by the upcoming XRISM \citep{2022arXiv220205399X} mission and its Resolve instrument.
The first high X-ray resolution spectra provided by its short lived predecessor, the SXS instrument onboard Hitomi \citep{2023arXiv230301642S}, in the direction of the Perseus cluster held very  promising perspectives on our ability to understand better the evolution and formation of galaxy clusters. There is hope that future data analysis methods will also be able to make full use of such spectra, allowing for combinations of spectral an spatial information in the cluster, and perhaps allow for the inclusion of the gravitational redshift as another useful probe to our understanding of these large structures.

\indent At the time of publishing this paper, the European Space Agency has sponsored a full reformulation of the Athena mission science case and specifications. We thus stress that the results of our study may have to be reconsidered according to the future new instrumental requirements of the Athena mission.

\begin{acknowledgements}
We are grateful to the anonymous referee for fruitful comments that helped improving this paper. AM, EP and NC acknowledge the support of CNRS/INSU and CNES. The following python packages have been used throughout this work : \texttt{astropy} \citep{astropy:2013, astropy:2018, astropy:2022}, \texttt{chainconsumer} \citep{Hinton2016}, \texttt{emcee} \citep{2013PASP..125..306F}, \texttt{matplotlib} \citep{Hunter:2007} and \texttt{cmasher} \citep{2020JOSS....5.2004V}. 

\end{acknowledgements}

\bibliographystyle{aa} 
\bibliography{biblio.bib}

\begin{appendix} 

\section{An analytical formula for $\delta_c$ in the case of a gNFW density profile} \label{AppedixA}

Let us recall the expression for the density contrast, $\delta_c$ : 
\begin{equation}
       \delta_c = \frac{M_{\delta}}{\int_0^{R_{\delta}} \frac{4 \pi r^2 \rho_{\text{crit}}(z)}{(r/r_s)^\gamma (1+r/r_s)^{3-\gamma}}}
\end{equation}
By using the expression for the gNFW density provided in \citet{2019arXiv190105615Z} and noting that :
\begin{equation}
    \int x^{2 - \gamma} (1+x)^{\gamma - 3} = \frac{x^{3-\gamma} \prescript{}{2}{F}_1(3-\gamma, 3-\gamma; 4-\gamma, -x)}{3-\gamma} + \mathrm{const.} 
\end{equation}
where $\prescript{}{2}{F}_1$ is the hyper-geometric function. We obtain the following formula for $\delta_c$ :

\begin{equation}
    \delta_c = \frac{M_{\delta}}{\frac{4 \pi r_s^3 \rho_{\mathrm{crit}}(z)}{3-\gamma} \prescript{}{2}{F}_1(3-\gamma, 3-\gamma ; 4-\gamma, - c_{\delta}) c_{200}^{3-\gamma}}
\end{equation}

\section{Simulations without background}
\label{appendixnobkg}

\begin{figure}[h!]
    \centering
    \begin{subfigure}[b]{0.4\textwidth}
        \centering
        \includegraphics[width=\textwidth]{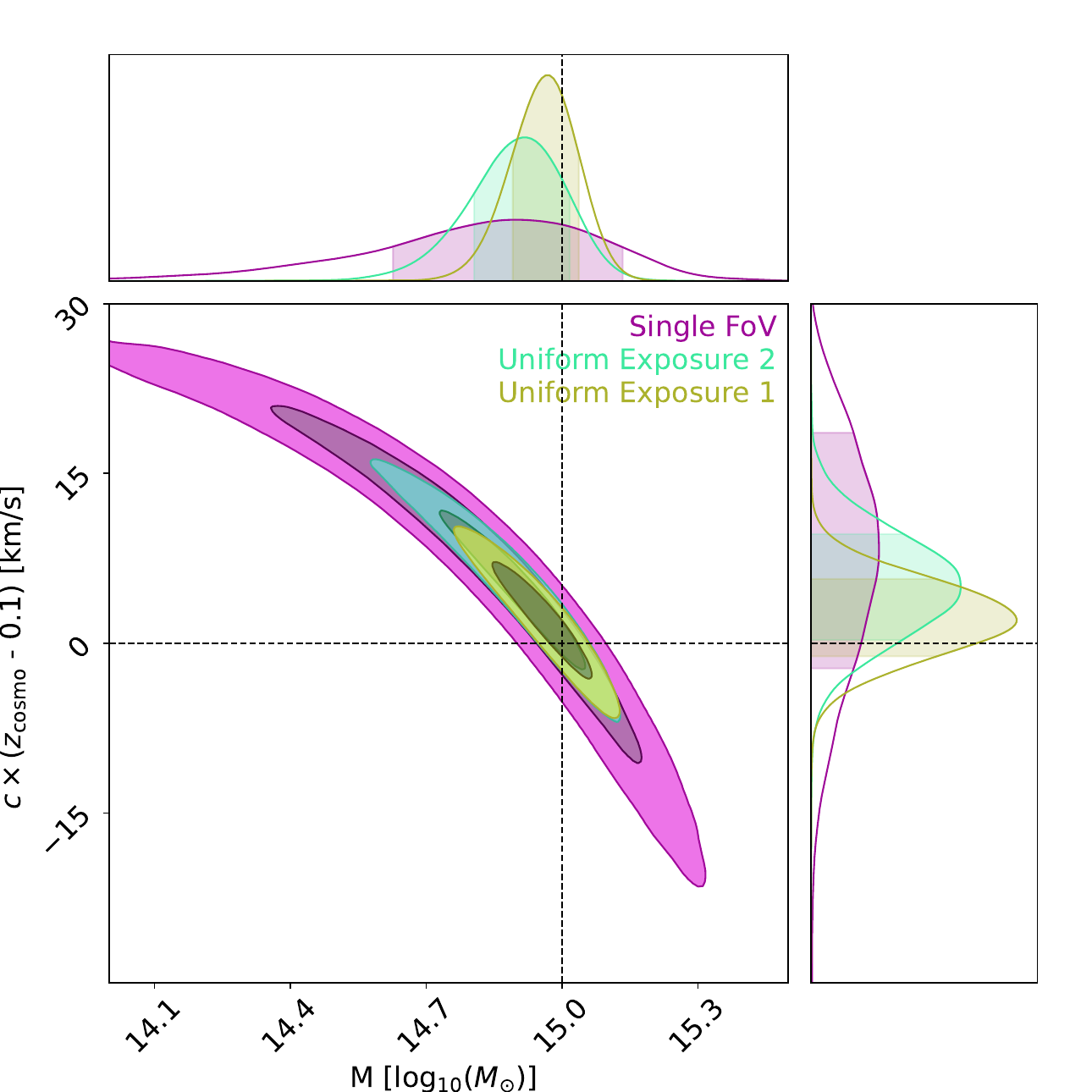}
        \label{uniformbkg}
    \end{subfigure}
    \vfill
    \begin{subfigure}[b]{0.4\textwidth}
        \centering
        \includegraphics[width=\textwidth]{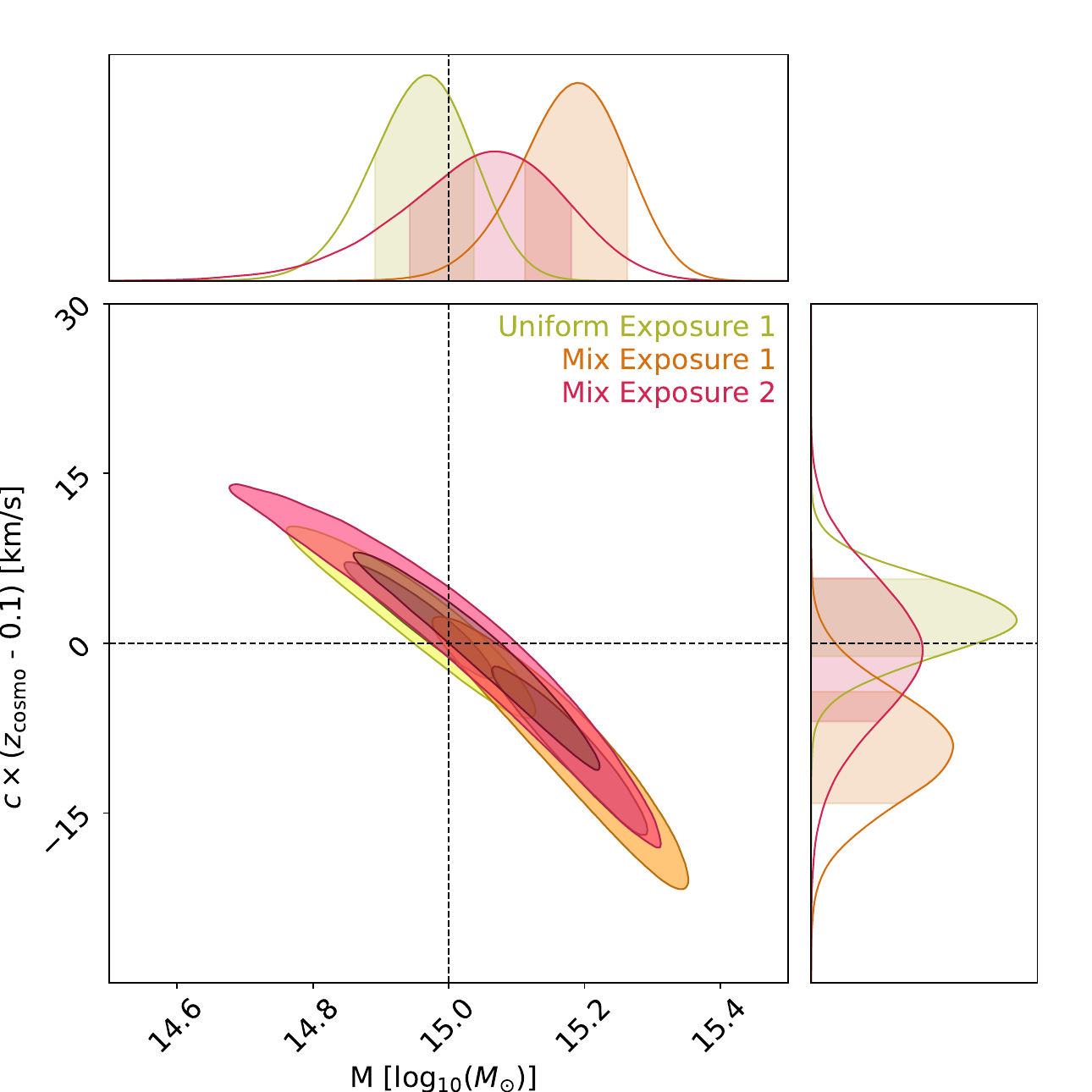}
        \label{mixnobkg}
    \end{subfigure}
    
    \caption{Corner plot in the mass and redshift plane and associated posterior distribution for both parameters. The different colors correspond to the different observing strategies listed in the legend and presented in Fig.~\ref{summarytable}. Top: Uniform exposures 1 and 2, compared with a single pointing observation. Lower: Mixed exposures 1 and 2, compared with a single pointing observation. The input reference values are plotted as dotted lines.}
    \label{cornerplotsnobkg}
\end{figure}

\indent The results presented in Sec. \ref{s:res} are obtained for simulations including astrophysical and instrumental background. We ran the simulations in the same setup, however without background, and ran the same analysis as presented in Sec. \ref{s:dataanalysis}. The results are shown in Fig.~\ref{cornerplotsnobkg}. For the observations with high exposure times, such as shown in the upper panel of Fig.~\ref{cornerplotsnobkg}, the background has little to no impact, and the constraints obtained over the parameters are identical to those in Fig.~\ref{cornerplots}. For observations with lower exposure times, however, the background is more predominant, and the difference between the cases with and without background is clearer. In particular,  mixed exposure 2 shows a smaller extent and is best centered on the real values. From these plots we conclude that the background can have a significant impact on the quality of the redshift fit and therefore on using the redshift profile as a probe for the gravitational redshift.


\clearpage
\pagebreak

\section{Fitting a line centroid with a binned observation}
\label{linefitting}

\begin{figure}[!h]
    \includegraphics[width = 9cm]{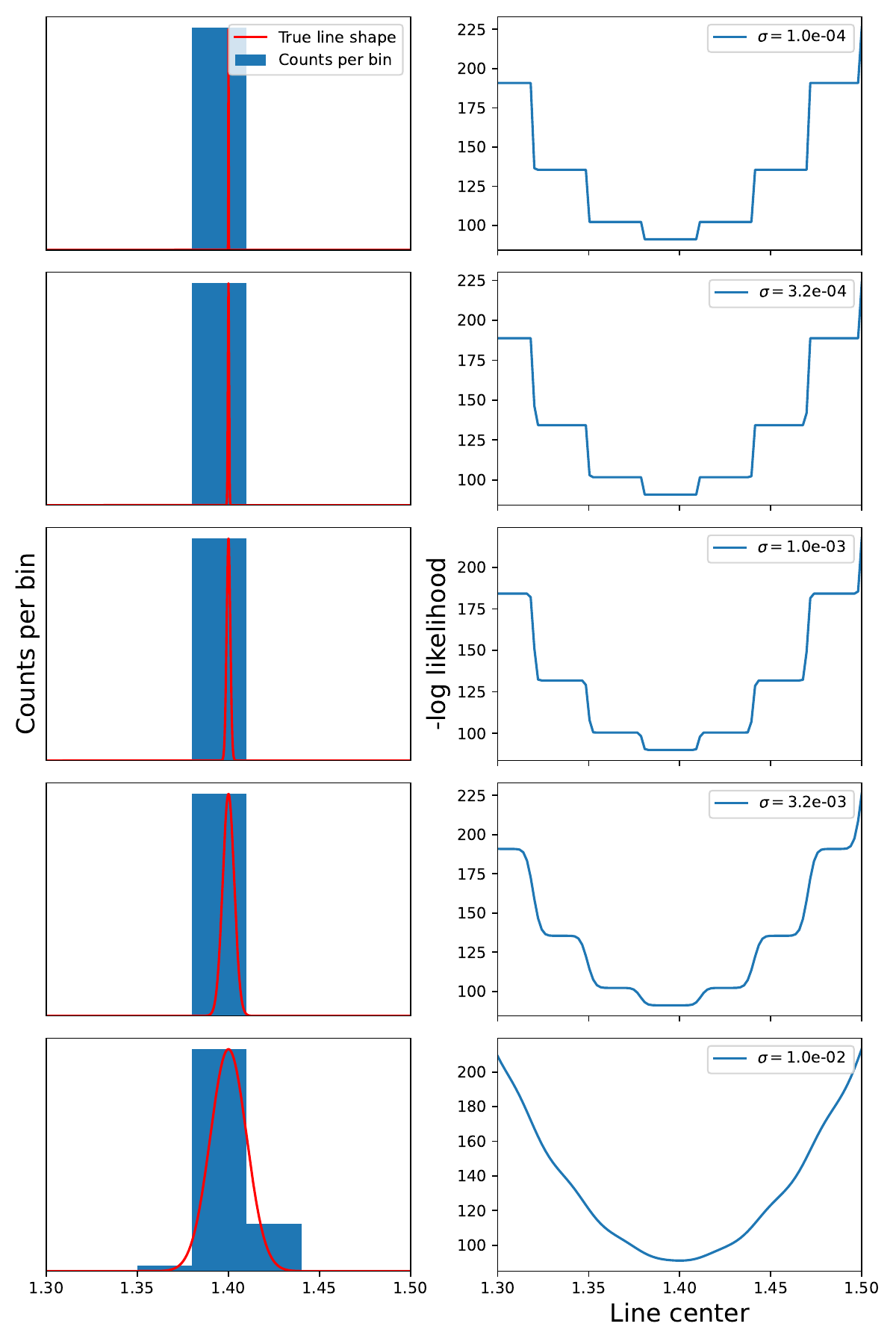}
   \caption{Illustration of the discretization occurring in the likelihood of a Gaussian model with respect to the line centroid when the line width is smaller than the binning of the data. Left: Counts per bin  as well as true line shape, as a function of energy, in the instrument for lines of different shapes, before instrumental broadening. The qualitative purpose of this plot allows for the use of arbitrary units on the vertical axis. Right: Likelihood of the Gaussian line model with respect to the line center, associated to the line observed in the left panel. }
   \label{likelihood}%
\end{figure}

\indent Here, we attempt to explain the discretization of the likelihood with respect to the  position of the emission line center (in the context of the observation of a line through any type of instrument that counts flux in discretized bins). This situation arises whenever the line width becomes equivalent or smaller than the bin width of the instrument measuring the flux. We generated line counts emitted by lines of different widths. For each line, we discretized these counts in bins, as is done with real data. Incoming photons are counted in energy bins to create the spectrum. Then, we broadened these fake line spectra by an arbitrary instrument response, of width slightly larger than the largest of the lines. This provides fake observations. 
For each line, we constructed a Gaussian line model and process it through the binning and instrument response in the same way as the "observed" counts are. Then, using Poisson statistics, we mapped out the likelihood with respect to the Gaussian line center, assuming the original line width is known. For lines with a width much lower than the instrument bin width, we show that the likelihood is not smooth anymore and has steps, the width of which approaches the bin width (see Figure \ref{likelihood}). Lines of small width with respect to the bin width do not affect the likelihood when their center is moved within the bins, thereby creating steps in the likelihood. In the context of fitting the line center, the likelihood minimum becomes quite ill-defined, as is   the associated uncertainties on the extracted parameter. 

\end{appendix}

\end{document}